\documentclass[letterpaper,twocolumn,10pt]{article}
\usepackage{usenix-2020-09}

\usepackage{tikz}
\usepackage{amsmath}
\usepackage{filecontents}

\usepackage{xlb}
\usepackage{multirow}
\usepackage{stfloats}
\usepackage{enumitem}
\usepackage{multirow}
\usepackage{array}
\usepackage{wasysym}
\usepackage{graphicx}
\usepackage{caption}
\usepackage{subcaption}
\usepackage{tikz}
\usepackage{graphics}
\usepackage{epsfig}
\usepackage{bm}
\usepackage{booktabs}
\usepackage{pifont}% http://ctan.org/pkg/pifont
\usepackage{xcolor, colortbl}
\usepackage{xspace}
\usepackage{listings}
\usepackage{boldline}
\usepackage[hang,flushmargin]{footmisc}
\usepackage{soul}
\usepackage{bbm}
\usepackage{tabularx}
\usepackage{minted}
\usepackage{amsmath, amsthm, amssymb}

\definecolor{codegreen}{rgb}{0,0.6,0}
\definecolor{codegray}{rgb}{0.5,0.5,0.5}
\definecolor{codepurple}{rgb}{0.58,0,0.82}
\definecolor{backcolour}{rgb}{0.95,0.95,0.92}
 
\lstdefinestyle{mystyle}{
    backgroundcolor=\color{backcolour},   
    commentstyle=\color{codegreen},
    keywordstyle=\color{magenta},
    numberstyle=\tiny\color{codegray},
    stringstyle=\color{codepurple},
    basicstyle=\ttfamily\footnotesize,
    breakatwhitespace=false,         
    breaklines=true,                 
    captionpos=b,                    
    keepspaces=true,                 
    numbers=left,                    
    numbersep=5pt,                  
    showspaces=false,                
    showstringspaces=false,
    showtabs=false,                  
    tabsize=2
}
\setlist{nolistsep}
\setlist[itemize]{itemsep=2pt, topsep=2pt}

\setstcolor{blue} 

\newcommand{\revaction}[1]{\textcolor{purple}{}}

\newcommand{\eat}[1]{}

\newcommand{\squishlist}{
 \begin{list}{${\bullet}$}
  { \setlength{\itemsep}{0pt}
     \setlength{\parsep}{1pt}
     \setlength{\topsep}{1pt}
     \setlength{\partopsep}{0pt}
     \setlength{\leftmargin}{1em}
     \setlength{\labelwidth}{0.5em}
     \setlength{\labelsep}{0.5em} } }

\newcommand{\squishend}{
  \end{list}  }
\newcounter{property}

\newcounter{packednmbr}

\AtBeginDocument{%
 \providecommand\BibTeX{{%
   \normalfont B\kern-0.5em{\scshape i\kern-0.25em b}\kern-0.8em\TeX}}}

\def\eg{{e.g., }}

\def\etc{etc.}

\begin{document}
%-------------------------------------------------------------------------------

\date{}

\title{\Large \bf \name: A High Performance Layer-7 Load Balancer for Microservices\\ using eBPF-based In-kernel Interposition}
\author{
{\rm Yuejie Wang}\\
Peking University
\and
{\rm Chenchen Shou}\\
Peking University
\and
{\rm Jiaxu Qian}\\
Peking University
\and
{\rm Guyue Liu}\\
Peking University
% {\rm Your N.\ Here}\\
% Your Institution
% \and
% {\rm Second Name}\\
% Second Institution
% copy the following lines to add more authors
% \and
% {\rm Name}\\
%Name Institution
} % end author

\maketitle

\begin{abstract}
    L7 load balancers are a fundamental building block in microservices as they enable fine-grained traffic distribution.
Compared to monolithic applications, microservices demand higher performance and stricter isolation from load balancers. 
This is due to the increased number of instances, longer service chains, and the necessity for co-location with services on the same host.
Traditional sidecar-based load balancers are ill-equipped to meet these demands, often resulting in significant performance degradation.

In this work,  we present \name, a novel architecture that reshapes L7 load balancers as in-kernel interposition operating on the socket layer.
We leverage eBPF to implement the core load balancing logic in the kernel, and address the connection management and state maintenance challenges through novel socket layer redirection and nested eBPF maps designs. 
\name\ eliminates the extra overhead of scheduling, communication, and data movement, resulting in a more lightweight, scalable, and efficient L7 load balancer architecture.
Compared to the widely used microservices load balancers (Istio and Cilium), over 50 microservice instances, XLB achieves up to 1.5x higher throughput and 60\% lower end-to-end latency.

\end{abstract}

\section{Introduction}
\label{s:intro}
The microservice architecture has gained widespread adoption in the construction of large-scale cloud applications~\cite{micro_example, micro_netflix, micro_uber, micro_airbnb}.
By decomposing a monolithic application into multiple manageable microservices, each service can be developed, deployed, and scaled independently.
This architectural transition leads to a substantial increase in the number of service instances and a surge in internal messages.
Load balancers (LB), acting as the pivotal link that distributes the majority, if not all, of these messages among services, have a significant impact on end-to-end performance.

Besides the increasing performance demand, the microservice architecture also introduces additional complexity to the operation of LB.
Our experience in managing large-scale cloud and microservice infrastructures has led us to identify three new unique challenges.
\begin{inparaenum}[1)]
\item The traditional Layer-4 (L4) LB approach proves too coarse-grained, necessitating the adoption of Layer-7 (L7) LB for more granular traffic management and enhanced application cooperation;
\item Unlike traditional architectures that host LB on separate servers, microservices deploy LB alongside services within the same host, introducing new performance and security challenges;
\item The invocation of multiple microservices to fulfill a single user request creates long service chains, demanding stringent performance requirements from each LB hop.
\end{inparaenum}

Given these new characteristics, the primary customer demands for LB are
\begin{inparaenum}[i)]
\item 
\textit{high performance} with minimal impact on application, and
\item \textit{strong isolation} between different applications and their respective LBs.
\end{inparaenum}
Unfortunately, existing sidecar-based approaches, such as Istio~\cite{istio} and Cilium~\cite{clium}, which deploy separate containers along with applications for load balancing (Figure~\ref{fig:arch}(a) and (b)), incurs significant overhead.
Our experiments (\S\ref{s:motiv}) show a 55.9\% reduction in throughput and a 2.27x increase in end-to-end latency under even light workloads.
To mitigate these issues, sidecar-less service mesh, such as Istio Ambient, delegates certain  Layer-4 networking functions to the kernel.
However, this approach cannot address HTTP-specific functionalities, which are conventionally managed by L7 LBs~\cite{sidecar_less,canal_sigcomm_24}.
Kernel-bypass techniques (Figure~\ref{fig:arch}(c)), such as DPDK~\cite{dpdk}, fail to shorten communication paths and require reimplementation of the network stack for L7 LB functionality.
Recent work~\cite{sr_osdi_23} implements LB as a library (Figure~\ref{fig:arch}(d) that runs inside applications, but it exposes the application to security risks~\cite{sure_socc_24} and requires application modification that loses the advantages of flexible deployment.
Thus, the library-based approach can only be deployed in internal trusted environments and cannot serve external customers in the public cloud~\cite{sr_osdi_23}.

In this work, we re-examine the L7 LB architecture with the aim of reducing overhead while maintaining strong isolation.
Our comprehensive analysis reveals that the traditional sidecar-based model inherently incurs unnecessary overhead.
Notably, we discovered that only a minor portion--about 20\%--of this overhead is attributable to essential LB functions such as protocol parsing and load balancing.
Thus, rather than operating as an independent scheduling entity, we advocate an L7 LB should be a synchronous interposition component on the data path.
Conceptually, LB acts as a \textit{logical extension} of the application, seamlessly integrated within the kernel.
Practically, we leverage eBPF~\cite{ebpf} for modular kernel modifications, messages interception, and executing LB operations without incurring additional overheads.

Employing eBPF to implement L7 LB poses a set of unique challenges, particularly due to the existing kernel lacking the necessary context for L7 traffic and connections.
The current socket subsystem is designed for basic data sending and receiving, but does not support advanced features such as connection splicing, which are crucial for load balancing.
Moreover, L7 LB requires maintaining various application states such as user configurations, protocol states, control information \etc.
These states, which are intricately linked, are traditionally managed with complex user space data structures.
Yet, eBPF, with its limited data structure support and stringent memory usage constraints, requires innovative approaches to accommodate these needs efficiently.

To address these challenges, we present \name, a novel in-kernel L7 LB.
Our key design decision is offloading LB to \textit{the socket layer} using eBPF programs.
Our insight is that the proposed interposition layer should be positioned as close to applications as possible.
This is contrary to recent data plane optimization approaches that move processing logic to lower layers, such as device drivers or hardware~\cite{nic_hotos_21, xdp_conext_18, xrp_osdi_22,bmc_nsdi_21}.
Although moving processing to lower layers can optimize I/O, it prolongs the execution path and increases complexity for L7 LB.
Our approach enables a modular extension and fast processing for in-kernel LB.
\name\ intercepts the application traffic with no extra cost from separate proxies, yet still, can operate on the complete application layer messages without the complicated interaction with networking stacks.

\begin{table}[]
  \centering
  \resizebox{0.8\columnwidth}{!}{
  \begin{tabular}{|l|l|l|l|}
    \hline
    System   & Istio  & Cilium    & \cellcolor[gray]{0.8}\name    \\ \hline
    Latency (us) & 1.433 & 1.197  & \cellcolor[gray]{0.8}0.63   \\ \hline
    Throughput (req/s) & 44652.8 & 53459.9  & \cellcolor[gray]{0.8}101449.6 \\ \hline
  \end{tabular}
  }
  \caption{\small Performance comparison of different L7 LB.
  }
  \label{tbl:moti}
\end{table}

To effectively expand the existing socket with the necessary L7 context, we have two key designs.
Firstly, we design an efficient \textit{socket message redirection} mechanism.
We enhance the current socket mechanism by associating a list of pools containing backend instances with a socket, bypassing connection handshake at the client side, and providing more flexible message redirection among sockets.
Secondly, we use \textit{nested eBPF maps} to build complex data structures to maintain states about flows, configurations, and services.
This makes \name\ compatible with existing control plane interfaces (\eg\ backend discovery and policy configuration) and supports all Envoy rules~\cite{envoy} related to load balancing used by Istio.
Deploying \name\ to serve a financial customer delivers a 41\% latency reduction and 30\% service density improvement.
To the best of our knowledge, \name\ is the first eBPF-based L7 LB production that securely serves external customers within a public cloud.

We make the following contributions :

\begin{packeditemize}
\item \textbf{New Architecture:}
A novel L7 microservices LB architecture that replaces sidecar-based model with in-kernel interposition in the socket layer using eBPF.
\item \textbf{Compatibility and Practicality:}
\name\ is compatible with the existing control and management plane and requires no application modifications, achieving a drop-in replacement of the current L7 microservices LB.
\item \textbf{Performance:} 
An exhaustive evaluation under various realistic workloads.
\name\ gets up to 1.5x higher throughput and 60\% lower average latency compared to the widely used LB (Istio and Cilium).
\end{packeditemize}

% \textit{This work does not raise any ethical issues.}

\section{Motivation and Background}
\label{s:motiv}
\subsection{Overhead Analysis of L7 LB}

\begin{figure*}[!t]
  \centering
  \includegraphics[width=0.95\textwidth]{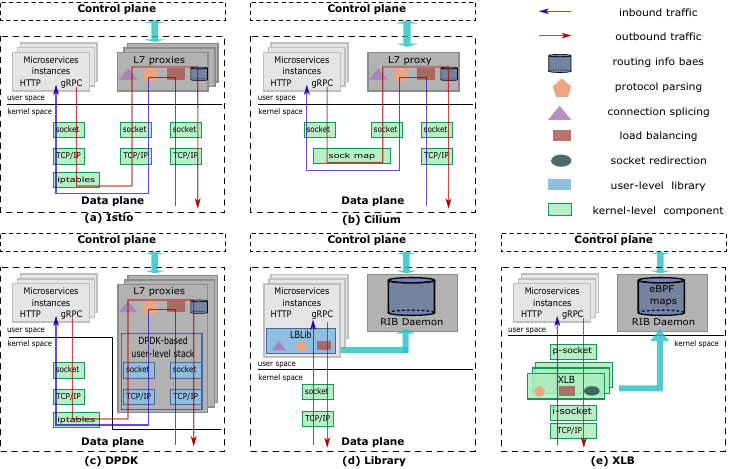}
  \caption{\small Different architectures of various L7 LB designs.
    (a) represents Istio, which involves redundant networking stack processing and has a per-service proxy.
    (b) illustrates Cilium that optimizes away networking stack and uses a global proxy for all services.
    (c) implements the sidecar with DPDK to bypass the kernel.
    (d) implements the LB as a library within the application process.
    (e) is \name, which offloads load-balancing logic into the kernel data path.
  }
  \label{fig:arch}
\end{figure*}

\head{Sidecar-based LB.}
There is a variety of L7 LB~\cite{haproxy, ngnix, envoy, istio, clium}.
In microservices, current dominant solutions, such as Istio~\cite{istio} and Cilium~\cite{clium}, use the \textit{sidecar-based} architecture shown in Figure~\ref{fig:arch}(a) and (b).
A sidecar is an independent container attached to a microservice instance. 
All traffic from and to the microservice is redirected into this sidecar that distributes the traffic among multiple instances at every hop.
Existing sidecar-less service mesh, such as Istio Ambient, cannot address L7 functionalities~\cite{sidecar_less}.
To quantify the overhead introduced by the current L7 LB sidecar, we conduct a micro-benchmark to measure the throughput and latency of two microservice instances with different LB (details in \S\ref{s:eval}).

Table~\ref{tbl:moti} shows the performance results of Istio, Cilium, and \name.
Cilium improves Istio by avoiding redundant networking stack in the kernel and using a global proxy for all services.
Despite these optimizations, Istio and Cilium incur substantial overhead.
Compared with \name, Istio (Cilium) results in a 55.9\% (47.3\%) reduction in throughput and 2.27x (1.9x) higher latency.
We next dissect the overhead to identify potential performance optimization opportunities.

\head{Performance overhead breakdown.}
We combine code instrumentation and the {\tt perf} flame graph to profile different components along the data path inside Istio.
We measure the latency between the sidecar receiving a request from the service and sending the request out.

Our profiling focuses on five key components:
\begin{inparaenum}[1)]
\item \textit{protocol parsing} which parses application layer messages;
\item \textit{load balancing} which distributes the load based on the policy (\eg\ weighted round robin, least load);
\item \textit{connection-splicing}  which tracks connections of both clients and servers and splices data streams among connections~\cite{splice_usits_99,splice_jounal_99,splice_atc_03};
\item \textit{socket processing} represents the communication cost between an application and the sidecar; and
\item \textit{kernel stack processing} is the part spent on the networking stack processing in the kernel, including both the application and the sidecar.
The rest of the overhead is accounted for by others, such as language (C++) overhead, event callback, and scheduling preemption.
\end{inparaenum}

Table~\ref{tbl:break} shows the time spent on each component.
From the results, we can see the overheads of essential load balancing functionalities, such as protocol parsing and load balancing algorithm only account for 20\%.
This shows that it is possible to eliminate a majority of L7 LB overhead without sacrificing key functionalities.
Two major sources of overhead are \textit{kernel protocol processing} and \textit{connection splicing} which in total account for more than 50\%, thus become our target of optimizations.
Diving deeper into these two components, we can further categorize the overheads into three types:

\textit{Duplicate protocol processing}:
Currently, both inbound and outbound traffic traverse the kernel TCP/IP stack three times (Figure~\ref{fig:arch}(a)), resulting in redundant processing.

\textit{System calls}:
Connection splicing involves numerous system calls, necessary for the creation, management, and termination of network connections, introducing more user-kernel switches and constant system call overhead.

\textit{Cross-process overheads}:
Running sidecar and application as separate processes, whether on the same core or different cores, incurs additional overhead from task placement, process communication, priority assignment, and event notification.
Even worse, these overheads are inevitably introduced to the system in one way or another.
For instance, if the LB and the service are placed on the same core, switching the context between them back and forth is necessary.
When they are placed on different cores, cross-core cache coherence then becomes unavoidable.
Another example is determining the scheduling priority of the LB.
Higher priority will block the application service, while lower priority will cause preemption from other unimportant tasks.

Next, we discuss possible approaches to eliminating the identified overheads.
The summary is shown in Table~\ref{tbl:lb_sum}.

\head{Kernel-bypass approaches.}
One method to mitigate kernel-related overhead is through kernel-bypass techniques, such as using DPDK~\cite{dpdk}.
By implementing the sidecar with DPDK (Figure~\ref{fig:arch}(c)), we can bypass the kernel, thereby eliminating the overhead associated with user-kernel switches and system calls.
However, this method bypasses the standard kernel protocol stack, necessitating a custom TCP/IP stack to enable L7 load balancing functionalities.
To optimize this custom stack further, particularly to minimize duplicate protocol processing by reusing results between the sidecar and the application, significant modifications to the application will be required.
Moreover, even when using DPDK, the sidecar and the application process still function as two distinct scheduling entities, leading to cross-process overhead.

\begin{table}
  \centering
\resizebox{0.75\columnwidth}{!}{
  \begin{tabular}{|c|c|c|} \hline
    \textbf{Component}&  \textbf{Sidecar-based}& \textbf{\name}\\ \hline
    Protocol-parsing&  4.5us (5.11\%)& ---\\ \hline
    Load Balancing&  13us (14.78\%)& ---\\ \hline
    Connection Splicing&  22us (25\%)& $\downarrow$\\ \hline
    Socket Processing&  3.83us (4.35\%)& $\downarrow$\\ \hline
    Kernel Protocol &  26.9us (30.62\%)& $\downarrow$ \\ \hline
    Others&  17.7us (20.12\%)& $\downarrow$\\ \hline
  \end{tabular}
  }
  \caption{\small Latency of different Istio data path components.}
  \label{tbl:break}
\end{table}

\head{Library-based approaches.}
Alternatively, implementing the LB as a library within the application  (Figure~\ref{fig:arch}(d)) can eliminate duplicate protocol processing and cross-process overhead~\cite{sr_osdi_23}.
Furthermore, this library-based approach comes with two new issues.
Firstly, it compromises the security model that isolates the application from the LB.
On the one hand, any vulnerabilities in the LB library could potentially expose the entire application to security risks, which is unacceptable to external customers.
On the other hand, malicious customers can change load balancing strategies to affect others.
Secondly, it requires application modifications and lost the advantages of easy deployment.
Thus, the library-based approach can only be deployed in internal trusted environments and cannot serve external customers in the public cloud~\cite{sr_osdi_23}.

\begin{table}
  \centering
  \resizebox{0.85\columnwidth}{!}{
  \begin{tabular}{|c|c|c|c|c|} \hline
    Overhead &  Sidecar &  Kernel bypass&  Library& \name \\ \hline
    Dup processing&  H&  H&  L& L\\ \hline
    System calls& H& L& H&L\\\hline \hline
    Cross-process &  H&  H&  L& L\\ \hline
    Isolation&  H&  H&  L& H\\ \hline
    Compatibility&  H&  L&  L& H\\ \hline
  \end{tabular}
  }
  \caption{\small Summary of LB optimization approaches.}
  \label{tbl:lb_sum}
\end{table}

\subsection{eBPF Background and Current Practice}

eBPF enhances the kernel's programmability, configurability, and flexibility by offloading customized logic to the kernel.
This approach avoids data movement and user-kernel switches, leading to considerable acceleration of user applications.
A typical eBPF program contains three types of elements:
\begin{inparaenum}[1)]
\item \textit{Kernel Hooks} serve as execution points where the program runs within the kernel;
\item \textit{eBPF Code} implements the logic of offloaded functions and requires verification for kernel safety.
It uses \textit{eBPF Maps} - simple kernel data structures - for persistent memory and communication with user applications; and
\item \textit{Helper Functions} support eBPF programs by providing additional functionalities that are not subject to verification, such as basic libraries or existing kernel functions, enhancing the programs' capabilities.
\end{inparaenum}

Currently, eBPF is used to optimize networking in two primary ways.
Firstly, eBPF can improve network measurements~\cite{sketh_sigreview_23,deepflow_sigcomm_23}, but these approaches cannot modify packet content which is necessary for LB.
The second type of work uses eBPF to achieve simple packet forwarding with L4 information~\cite{iptable_signcom_19,spright_sigcom_22,xdp_conext_18}.
For example, SPRIGHT~\cite{spright_sigcom_22} forwards packets within a service chain in the same node.
These works cannot interpret application message semantics and states to achieve richer routing and load-balancing logic.

\section{\name\ Overview}
\label{s:over}
\subsection{Design Goals}
\label{ss:goal}

We desire to offload L7 LB into the kernel, where \name\ is safely and efficiently integrated as part of applications.
\name\ needs to satisfy the following goals to be efficient and practical.

\head{Near-zero unnecessary overhead.}
\name\ should eliminate unnecessary data movement, process scheduling, and redundant processing to avoid overhead beyond its essential functionalities (\eg\ protocol parsing and load balancing).

\head{Service isolation and security.}
\name\ must effectively maintain isolation between multiple services running on the same host as well as protect itself from malicious users.

\head{Operational compatibility.}
For \name\ to be seamlessly integrated on a production scale, it must be compatible with existing control and management systems and transparent to users, requiring minimal effort for adoption.

\subsection{Design Challenges and Insights}
\label{ss:insight}

Modifying the kernel in a blind and intrusive manner to implement L7 LB not only falls short of meeting the above goals but can also increase risks and maintenance costs.
While eBPF allows for modular and non-intrusive modifications to kernel functions, the implementation of L7 LB using eBPF poses unique challenges, primarily because the existing kernel lacks the necessary context for L7 traffic and connections.

\head{Challenge 1: insufficient and inflexible connection management in the kernel.}
L7 LB frequently migrates and splice application flows among multiple backend connections.
Traditional proxy implementations maintain connection pools and invoke ordinary socket operations on individual sockets, which incurs extra system call overhead.
The kernel is not aware that these sockets work together for the same application flow, thus \name\ has the opportunity to integrate the management operations into the kernel to eliminate extra overhead.
However, the current socket subsystem in the kernel is not designed to support LB, lacking the appropriate data structures and workflow for load balancing.

\head{Challenge 2: complex application layer states maintenance in the kernel.}
L7 LB maintains a large number of internal application states, including user configurations, control information, traffic statistics, protocol states, and flow mappings.
These states are interrelated and coupled, requiring complex data structures to represent and maintain.
While \name\ operates as an independent kernel layer, it still faces the challenge of managing states in applications and LB.
This aspect of state management in \name\ is particularly challenging, since limited data structures are available and the kernel has stricter requirements in terms of memory consumption and safety.

\name\ adopts a novel approach driven by two fundamental rationales:
\begin{inparaenum}[1)]
\item operating L7 LB as independent scheduling entities (such as processes, containers) introduces unavoidable cross-process overhead;
\item running the LB in user space leads to either costly user-kernel switches or duplicated processing that could be efficiently handled within the kernel.
\end{inparaenum}

\head{Observation 1: LB should be a logical extension of the application.}
XLB conceptualizes the LB not as a separate entity, but as a \textit{logical extension} of the application, managed within the kernel's datapath.
As illustrated in Figure~\ref{fig:arch}(e), \name\ allows applications to use sockets for communication as normal, while the load-balancing logic seamlessly integrates into the kernel.
This integration significantly reduces the overheads associated with cross-process communication and frequent transitions between user and kernel space.

\head{Observation 2: offloading LB to the socket layer with eBPF programs is the key.}
To effectively address these challenges using eBPF, the foremost design decision is identifying the most suitable the eBPF hook location.
This decision is influenced by the depth of the fast code path and the availability of kernel context around the hook points.
Recent work prefers hooking eBPF programs at lower levels, such as device drivers or hardware~\cite{nic_hotos_21, xdp_conext_18, xrp_osdi_22,bmc_nsdi_21} to optimize I/O performance because of the immediate I/O data access.
However, this approach results in significant overhead for functions operating at the application layer since user data are generated from applications instead of I/O devices. 

In-kernel L7 load balancing should be able to manipulate application-level messages efficiently while trying to avoid involving other kernel functionality to minimize intrusive modification and extra overhead.
So we choose to offload LB exclusively to the \textit{socket layer}, which is closest to the application layer.
This design decision brings several benefits:
\begin{itemize}[leftmargin=*]
\item \textit{Complete message contents.}
Operating at the socket layer can intercept the complete application messages without complicated packet reassembly in the TCP layer.
\item \textit{Fast message process.}
Some unnecessary overhead in the networking stack or other kernel functionality at a lower level can be avoided in case sockets are able to process the message immediately, such as no route drop.
\item \textit{Flexible message rewriting.}
Modifying messages at the socket layer is more flexible, because lower levels, such as device drivers, lack enough context to perform these modifications and incurs repeated operations.
\item \textit{Modular extension.}
Socket layer offloading decouples \name\ from the networking stack with better modularity.
In this way, \name\ does not need to maintain TCP states, but can still cooperate with other mechanisms in the lower levels, such as iptable and XDP~\cite{xdp_conext_18}.
\end{itemize}

\subsection{\name\ Architecture Overview}
\label{ss:overview}

\head{Design 1: Enhanced Socket Redirection (\S\ref{ss:socket}).}
To maintain the relay relationship effectively, \name\ introduces two new socket types, expanding the socket data structure with a connection pool and a request map.
\name\ extends the socket operation flows using eBPF programs, which involves breaking down operations into smaller, more manageable steps crossing the connected socket pair.
This approach ensures more efficient and precise socket operations.

\head{Design 2: Efficient Application State Management (\S\ref{ss:state}).}
It is important for \name\ to be compatible with existing LB control and management operations, thus \name\ cannot simplify or discard some application states.
To address this, we utilize nested eBPF maps to build complex data structures, while carefully optimizing their memory usage to ensure efficiency and effectiveness in state management.

Figure~\ref{fig:arch}(e) shows the overall architecture of \name\ incorporating these key designs.
The L7 LB is a collection of eBPF programs of different functionalities hooked into the socket layer to intercept service messages.
In addition, service messages are also inspected with the application context of each request.
\name\ forwards different requests to appropriate backends based on load-balancing algorithms.
Flow states and backend information are stored in eBPF maps, ensuring configurations are compatible with the existing control plane.

\section{\name\ Design}
\label{s:design}
Similar to the existing LB, the main workflow in \name\ includes connection establishment, load balancing, and message forwarding.
The client's TCP connection is bypassed because no backend can be determined at this point without L7 messages.
During load balancing, \name\ determines the destination service with content-based routing and chooses one instance according to the load balancing algorithm.
Message forwarding is performed between two sockets in the kernel using the socket relay mechanism (\S\ref{ss:socket}).

\subsection{Socket Proxy and Redirection}
\label{ss:socket}

\begin{figure}[t]
  \centering
    \includegraphics[width=0.99\columnwidth]{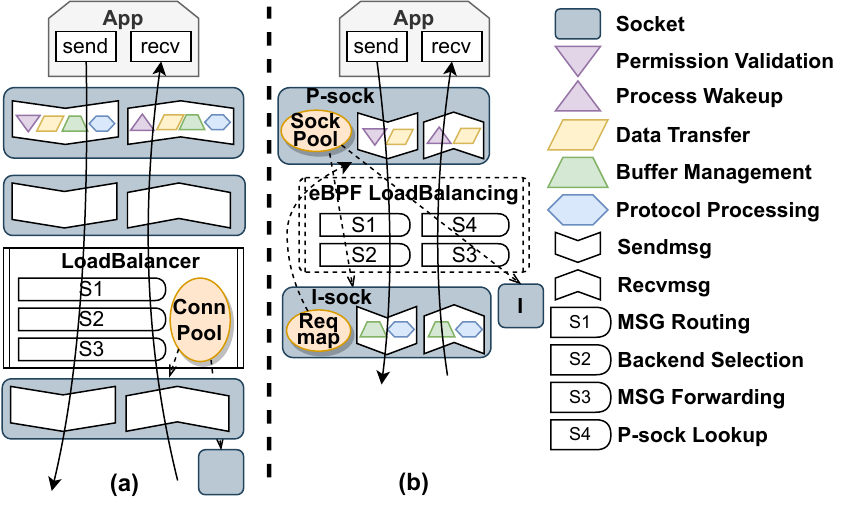}
  \caption{\small Comparison of socket subsystem between the current kernel and \name.}
  \label{fig:sock_comp}
\end{figure}

Redirecting data between connections is the basic ability of load balancing.
Figure~\ref{fig:sock_comp} compares \name\ socket enhancements with conventional socket usage in existing LB.
As shown in Figure~\ref{fig:sock_comp}(a), the current LB involves multiple independent sockets, incurring redundant computation and message redirection.
Some Linux improvements (\eg\ sockmap in Figure~\ref{fig:arch}(b)) are usually used together with L4 information but are still not sufficient to support L7 processing.
Figure~\ref{fig:sock_comp}(b) illustrates the improved socket subsystem in \name.

\head{New socket type.}
\name\ extends Linux sockets with two extra types: proxy socket (\ps) and instance socket (\is).
A \ps\ is created by the client during connection establishment, holding the core logic of load balancing.
\is s are created by \name\ to communicate with servers, which are pre-established with the backends before any client connections.
\name\ manages message redirection between these two types of sockets, and this socket relay is much more lightweight than connection tracking in existing proxies.

\head{Expanded data structure with connection pool.}
A \ps\ is associated with an \is\ connection pool.
Choosing the granularity of connection pools needs to consider various trade-offs.
A larger capacity will cause more contention from concurrent connections, while a smaller capacity brings too many pools and constant overhead.
So currently, \name\ shares a connection pool within each {\tt cgroup}, therefore preserving the resource isolation guarantee in the container subsystem.
Items in the connection pool save the tuple of IP and port, which is used to look up the corresponding \is.

\head{Expanded data structure with request map.}
There is more involved when multiple requests from different \ps\ are distributed to the same \is.
If the L7 protocol requires in-order response (\eg\ HTTP/1.1), \name\ maintains only one outstanding request and holds other requests until the previous response is received.
If the L7 protocol allows concurrent requests (\eg\ HTTP/2.0), \name\ has to correctly return a response to its request.
Fortunately, L7 protocols always include some request identifiers in such cases (\eg\ stream id of HTTP/2.0).
Thus, \name\ attaches a request map to each \is, internally allocates request identifiers, and maps the internal identifier to the original one.

\head{eBPF interposition.}
Similar to existing LB, \name\ first decides the destination service and then selects one instance from the service cluster.
But \name\ implements all these logic with eBPF and executes them synchronously in the kernel.
\name\ combines message contents with IP addresses to locate the actual service,
then matches requests with routing rules sequentially, and the last matched rule resolves the destination service.
To select an instance from the service cluster, \name\ supports conventional loading balancing algorithms, such as round-robin, random, and the least request.

\head{Socket relay.}
The \ps\ relays L7 messages to the \is\ determined above.
In contrast to sockmap used by Cilium which forwards outgoing requests to the RX queue of the user-level proxy's socket (Figure~\ref{fig:arch}(b)), \name\ relays them to the TX queue of the \is.
Similarly, incoming messages are redirected from the \is's RX queue to the \ps's RX queue.
The \is\ modifies requests and responses using proper identifiers.
More implementation details of kernel modifications to the socket will be discussed in \S\ref{ss:ko}.

\begin{figure*}[!t]
  \centering
  \includegraphics[width=7.1in]{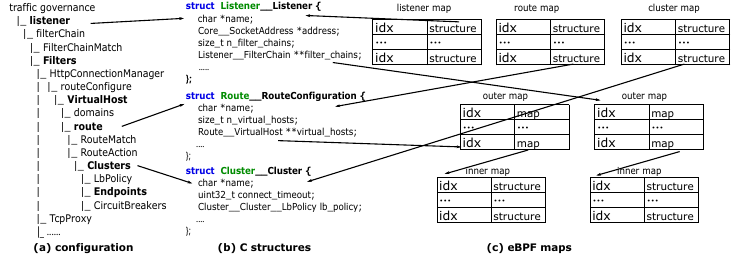}
  \caption{\small A simplified example of converting Envoy configurations to C structures and eBPF maps.
  Configurations are in a tree hierarchy, and eBPF maps are organized using map-in-map structures.
  }
  \label{fig:impl}
\end{figure*}

\subsection{State Management}
\label{ss:state}

\name\ maintains three types of states with eBPF maps.
The first is user configurations, including service management, routing rules, and cluster setup.
The control plane distributes and updates these states to \name.
They are read-only on the data path and modified by user user-level control daemon.
Figure~\ref{fig:impl} gives a concrete example of configurations and their associated eBPF maps.
Another is load-balancing states.
Different algorithms may require different sets of states.
For example, to support the least request policy, \name\ tracks the number of requests forwarded to each connection or instance.
While these states may be updated concurrently, the eBPF map handles synchronization internally.
The third type of state collects traffic metrics for each flow, such as TX/RX bytes, request counts, and no\_route\_match.
As they bind to each flow, updating them causes no concurrency issues.

\head{eBPF map conversion.}
Envoy configurations are designed as tree hierarchies (Figure~\ref{fig:impl}(a)).
Thus, \name\ needs to define C structures and maintain the hierarchy in eBPF maps accordingly while minimizing the memory cost.
eBPF maps do not support direct pointers and require key-value style reference to represent a hierarchical relationship in the tree.
Additionally, extra metadata and statically allocated map entries increase memory consumption.
We leverage the {\tt map-in-map} to create references between maps.
We replace pointers with map ids and indexes.
Outer maps translate a map id to an inner map saving actual values.
Some global named maps serve as the root entrance.
Figure~\ref{fig:impl}(c) demonstrates the relationships among different maps.
However, verification requires statically defining and allocating eBPF maps, which may cause large memory consumption.
We make two optimizations to compress the size of the value and the map capacity.
\begin{inparaenum}[1)]
\item These maps save the actual structures, thus a large structure (\eg\ a structure containing an array or nested structures) takes lots of space.
We lift these nested members to separate maps and use {\tt map-in-map} to reference them.
As a result, the size of a nested member is reduced to the fixed size of map indexes.
\item The map capacity is statically defined and determines the total map size.
To avoid wasting memory on a huge map, we limit the map capacity to 10K.
All these eBPF-related operations are implemented using {\tt ebpf-go} library~\cite{ebpf_go}.
\end{inparaenum}

\head{Delta refresh.}
When updating configurations, \name\ only deals with the changed items.
However, to maintain the tree structure, we have to carefully modify these interconnected eBPF maps.
To add an item, we follow the bottom-up order.
We add a new item in the lowest level map first and then modify its parent all the way up to the top level configuration.
Deleting items uses a top-down approach instead.
In this way, configuration updates do not disturb the kernel data path accessing configuration rules.

\subsection{Deployment Practice}
\label{ss:deploy}

\head{Customer usecase.}
We deploy \name\ to serve a banking customer whose service processes online payments to support mobile banking transactions.
The payment processing system is constructed using micro-services and consists of dozens of services.
The customer has extremely high latency and security requirements, and each transaction of online payment must be finished with guaranteed low latency and high security.
Therefore, neither sidecar-based nor library-based L7 LBs satisfy the requirement, \name\ is the most appropriate solution available for commercial customers.

\head{\name\ usage and effect.}
Our customer deploys \name\ in a cluster comprising over 100 nodes, and each node is an ARM server with Kunpeng-920 CPU.
There are tens of service instances running on each node.
Since \name\ offloads load balancing into the kernel to avoid extra packet redirection between sidecars, \name\ optimizes the transaction processing latency by 41.8\%.
Thanks to the eliminated sidecar and its associated CPU overhead, \name\ reduces the CPU consumption by an order of magnitude.
Moreover, the saved CPU and resource can be allocated to execute more instances within each node to enhance the elasticity and scalability, which is crucial for bank payment processing to support burst transactions in holiday sale (\eg\ black friday).
\name\ delivers a 30\% improvement of the service instance density to the customer.

\subsection{Discussion and Summary}
\label{ss:diss}

\head{Development.}
One concern about \name\ is its development effort due to eBPF programming.
Based on our experience, the development effort is acceptable because it is not hard for senior kernel developers to master common tricks in eBPF programming, such as tail calls and bounded memory access.
Moreover, eBPF-based solution is more stable after going online.
In contrast, our other services that use kernel-bypass and DPDK, occasionally crash due to undiscovered bugs, causing more maintenance costs and customer complaints.

\head{Practicability.}
The viability of \name\ in production stems from three design advantages.
\begin{inparaenum}[1)]
\item It is compatible with control plane, ensuring seamless integration into existing operational workflows. 
\item It requires no modifications to user applications.
\item Kernel modifications are implemented via eBPF and kernel modules, which is common practice employed by OS vendors.
\end{inparaenum}

\head{Extensibility.}
Some computation-intensive operations (\eg\ TLS handshake, authentication, and authorization) are difficult to implement with eBPF.
They also dominate the performance, thus offloading them into the kernel has little performance benefit.
Improving \name\ for better flexibility and extensibility is ongoing work. % in our project.

\head{Summary.}
We summarize how \name\ satisfies the design goals listed in \S\ref{ss:goal}.
By moving the load balancing into the kernel, \name\ avoids irrelevant performance costs of scheduling, communication, and data transfer.
Furthermore, in-kernel processing protects \name\ from malicious user services.
\name\ runs synchronously as the service execution, thus it avoids timing interference from other concurrent services.
As resource allocation adheres to {\tt cgroup}, \name\ also minimizes resource contention.
Lastly, \name\ focuses on the data plane and converts control plane configurations into eBPF maps (\S\ref{ss:state}), thus \name\ is compatible with the current microservice configurations and transparent to end users.

\section{\name\ Implementation}
\label{s:impl}
The \name\ implementation consists of three main parts: the modified socket mechanism inside the kernel, eBPF programs with auxiliary helper functions, and a control plane daemon.
Our current version has around 26K lines of code in total, including 3000 lines for kernel modification, 8000 lines for eBPF-related code, and 15K lines for the control plane.

To be compatible with the Envoy control plane, \name\ has a daemon running on each node.
This daemon is implemented in Go, and receives control information (\eg\ configurations and instance states) from cluster management using gRPC.
As shown in Figure~\ref{fig:impl}(b), \name\ needs to define C structures and maintain the configuration hierarchy in eBPF maps accordingly.
All data are transmitted in protobuf, which is then converted to C structures in Figure~\ref{fig:impl}(b) using {\tt protoc-c}.

\subsection{Kernel Modifications}
\label{ss:ko}

\name\ augments the socket subsystem with \ps\ and \is.
The service IP for creating a \ps\ is usually a virtual IP, so multiple services can sit behind the same IP.
For example, in canary testing, multiple versions of a service share the same IP, but different users are routed to different versions (\eg\ internal users go to the newer version).
\name\ matches a request with each of the routing rules organized as a chain by the control plane.
First, \name\ parses the application protocol to extract and save target fields (\eg\ URL or username).
A field with its key specified in the rules is retrieved and compared with the user-configured value.

The connect and send path in \ps\ have an eBPF hook.
During a connect system call, \name\ uses the destination IP and port to check whether this connection needs L7 load balancing based on the user configuration.
If not, \name\ sets a flag to bypass the rest of the procedure.
If the L7 load balancing is desired, \name\ invokes the modified {\tt sendmsg} operation of \ps,
which first triggers the eBPF hook to determine the destination \is, and then directly calls the \is's {\tt sendmsg} to relay the message.

\name\ replaces both send and receive paths in \is.
On the send path, \name\ maps requests to \ps\ and potentially rewrites the request id (\S\ref{ss:socket}).
For the receive path, we modify the {\tt sk\_data\_ready} operation called by {\tt tcp\_data\_ready} from the networking stack.
The original implementation notifies the process of waiting for a message.
For example, a process may block on {\tt epoll} or {\tt receive}.
However, \is\ needs to skip this notification and passes the message to the corresponding \ps\ identified by the request id.
\ps\ completes the final receiving process.

\subsection{eBPF Programs}
\label{ss:ebpf_impl}

We split the core functional logic of \name\ into multiple small eBPF programs based on socket operations and the hierarchy of configuration rules as shown in Figure~\ref{fig:impl}(a), and chain them via tail call.
The key challenges of eBPF programming are passing verification under constraints of bounded memory and limited instruction count.

\head{eBPF verification.}
There are three considerations about the verification.
\begin{inparaenum}[(1)]
\item Bounded loop.
Infinite loops are forbidden, as they can stuck the kernel.
Fortunately, all the configurable parameters have constant predefined maximal values, \eg\ the number of instances, routing rules and connections.
\item Memory safety.
Similar to the XDP program~\cite{xdp_conext_18}, to ensure memory safety at verification, we pass both the start and the end address of the accessed buffer as eBPF arguments, and explicitly check every memory access falls within the specified range.
\item Unsupported functions.
Due to the limited capability, eBPF cannot support arbitrary functions and requires helper functions.
Helper functions will not be verified, thus they are more powerful, but for the same reason, we have to choose the granularity of helper functions carefully.
\end{inparaenum}

\begin{figure}[!t]
\begin{lstlisting}[numbers=left,numbersep=3pt,xleftmargin=2em,framexrightmargin=-0.9em,framexleftmargin=1.1em,showlines=true,numberblanklines=false,style=eulerc,otherkeywords={bpf_get_msg_element,bpf_strcmp,tail_call,bool}]
int filter_manager(ctx_buff_t *ctx)
{
    for (unsigned int i = 0; i < FILTER_MAX_NUM; i++) {
        // retrieve pointer value from eBPF map
        filter = get_ptr_val(ctx->filters + i); 
        if (filter->config_type == FILTER_CONFIG_HTTP) {
            // extract value from L7 packet header
            if (bpf_get_msg_element(proto)== PROTO_HTTP_1_1) {
                match = 1; break;
            }
        }
    }
    // invoke next layer in the configuration hierarchy 
    if (match) tail_call(route_manager); 
}
    
int route_manager(ctx_buff_t *ctx)
{
    for (unsigned int i = 0; i < ROUTE_MAX_NUM; i++) {
        route = get_ptr_val(ctx->routes + i);
        ele = route->match_filed;
        exp_val = route->match_value;
        msg_val = bpf_get_msg_element(ele);
        // compare packet filed with route configuration
        if (bpf_strcmp(exp_val, msg_val) == 0) {
            match = 1; break;
        }
    }
    if (match) tail_call(load_balancer);
}
\end{lstlisting}
\caption{\small Simplified eBPF code for filter and route layer in Envoy configuration hierarchies (\S\ref{ss:state}).}
\label{fig:bpf_func}
\end{figure}

\head{eBPF functions.}
Figure~\ref{fig:bpf_func} gives the eBPF code for filter and route layer in Envoy configuration hierarchies (\S\ref{ss:state}) as examples.
Each hierarchy layer has similar processing logic.
We iterate all user configured items in a bounded loop (line 3) and check if the packet matches the configuration (line 6).
If matched, we invoke the next hierarchy layer using tail-calls.
Function {\tt filter\_manager} shows a L7 filter which matches the HTTP/1.1 protocol.
Function {\tt route\_manager} is a simplified content-based routing procedure.
User-configured routes include matching filed and expected value.
Line 23 retrieves the actual value, compares it against the expected value, and decides the destination service host.

\head{Helper functions.}
We either implement helper functions according to existing ones in the kernel or use them to provide very basic library functions.
\begin{inparaenum}[1)]
\item Message modification.
\name\ needs to allocate new memory when expanding messages (\eg\ insert headers).
We utilize the existing helper {\tt bpf\_msg\_push\_data} in {\tt sockmap} for dynamic memory allocation.
\item HTTP parsing.
While parsing HTTP protocol is complex, we find it ultimately relies on basic string operations, which cannot be verified due to non-constant loop bounds.
Thus, we wrap primary string operations (\eg\ {\tt strcmp} and {\tt strstr}) in helpers.
\item Regular expressions.
Content-based routing matches routing rules using regular expressions, which cannot be verified.
Fortunately, open source community provides practical kernel helper functions for processing regular expressions~\cite{regex_netdev}.
We reuse their implementation~\cite{regex_github}.
\end{inparaenum}

\begin{figure*}[!t]
  \centering
  \includegraphics[width=6.7in]{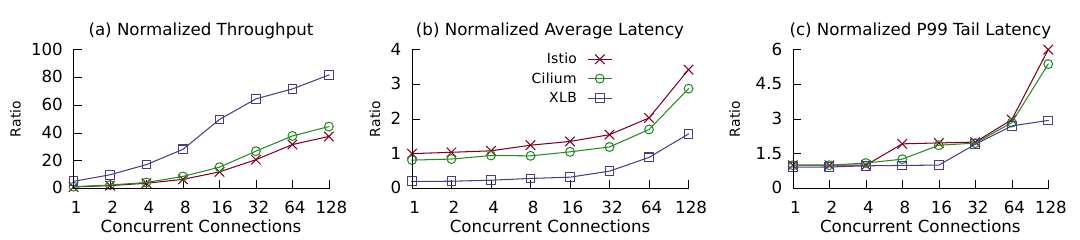}
  \caption{\small Micro-benchmark results of various number of concurrent connections with zero size HTTP payload.
  All data are normalized to Istio single connection result.
    }
  \label{fig:micro_conn}
\end{figure*}

\section{Evaluation}
\label{s:eval}
\begin{figure*}[!t]
  \centering
  \includegraphics[width=6.7in]{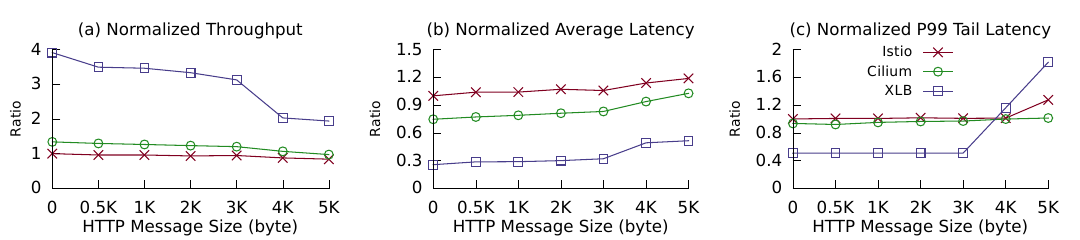}
  \caption{\small Micro-benchmark results of increasing message (HTTP payload) size with 200 connections.
    All data are normalized to Istio zero size message result.
  }
  \label{fig:micro_size}
\end{figure*}

Our experiments use x86 servers with two 20-core NUMA nodes (80 hyperthreads in total).
Some experiments use virtual machines with 16 cores of the same CPU. % and 32G RAM.
All the nodes are in the production environment and share networking with other internal production traffic.
All experiments use openEuler 22.03~\cite{oe} operating system based on Linux 5.10 kernel.
We use Fortio v1.50.0 to issue HTTP/1.1 requests using the persistent connection in a closed loop.
Service instances run in Kubernetes pods.
We compare \name\ with industry LB, Istio v1.15.3 and Cilium v1.12.2.
We configure Istio to use per-core worker, and Cilium uses its internal fixed configuration of proxy workers.
As \name\ supports the Envoy control plane, all studied systems use the same service and Kubernetes configurations, and they all use the least requests load balancing policy.

\subsection{Micro-benchmarks}
\label{ss:micro}

The micro-benchmark has one client and one server service, and the two services run on separate machines.
The server includes two instance pods and configures a single URL prefix matching routing rule.
We study different parameters and report the aggregate throughput across all requests, the average latency, and the 99th percentile tail latency.

\head{Concurrent connections.}
Figure~\ref{fig:micro_conn} reports the normalized result under different numbers of concurrent connections.
Each client thread creates one connection, thus the connection count is also the number of threads.
Since this experiment does not have an HTTP payload, the throughput of all systems gets higher with more connections.
Cilium is better than Istio because Cilium avoids unnecessary networking stack overhead as shown in Figure~\ref{fig:arch}(b).
However, \name\ outperforms both of them as it completely avoids the user-level proxy process.
For example, with 128 connections, \name\ has 2.18x and 1.83x higher throughput than Istio and Cilium, respectively.
With more concurrency, both the average and tail latency increase, but \name\ always has the lowest latency because of the proxy-less data path.
As shown in Figure~\ref{fig:micro_conn}(c), \name\ considerably reduces the tail latency by eliminating unpredictable system overhead such as scheduling and cache coherence.
In the case of 128 connections, \name\ reduces the tail latency by 51\% (45\%) compared to Istio (Cilium).

\head{Message size.}
Figure~\ref{fig:micro_size} depicts the performance with various message sizes.
Similar to the above experiment, the efficiency of \name\ is seen from the improved throughput, reduced average latency, and smaller tail latency.
\name\ again achieves the highest throughput of all message sizes.
After the size exceeds 4K, \name\ saturates the NIC, thus its throughput begins to drop (Figure~\ref{fig:micro_size}(a)),
while Istio and Cilium never saturate the NIC due to low throughput.
Ideally, the message size has less impact on the latency since most processing logic operates on message headers, not payload.
However, the proxy model introduces more data copying, thus resulting in more latency increases.
For example, the tail latency of Istio (Cilium) increases by 27.5\% (8.3\%), while \name\ only increases by 0.24\% before saturating the NIC.
On the other hand, \name\ has a large growth in the tail latency after it saturates the NIC due to much queuing delay as shown in Figure~\ref{fig:micro_size}(c).

\begin{figure*}[!t]
  \centering
  \includegraphics[width=6.7in]{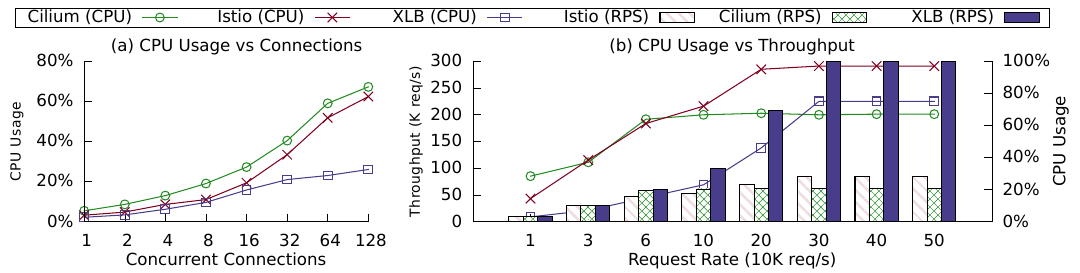}
  \caption{\small Aggregated CPU utilization of all services.
    (a) shows CPU usage of an increasing number of connections.
    (b) shows CPU usage under fixed throughput with 200 connections.
    }
  \label{fig:cpu_util}
\end{figure*}

\head{CPU usage.}
We measure the CPU utilization under different workloads.
Figure~\ref{fig:cpu_util}(a) shows that the CPU utilization grows with the increased connection count when the CPU is not fully utilized.
\name\ achieves the highest throughput and surprisingly uses less CPU.
For instance, with 128 connections, \name\ uses 58.3\% (61.3\% ) less CPU than Istio (Cilium).
This confirms that a majority of overhead in the existing L7 LB can be eliminated with the in-kernel load balancing.
We further increase the connection count to 200 and fix the message rate, then plot the throughput and CPU utilization in Figure~\ref{fig:cpu_util}(b).
When the rate is slower than 60K req/s, all systems achieve the configured rate, but \name\ requires the least CPU, because \name\ avoids wasting CPU in unnecessary communication and data movement.
\name\ continues to get higher throughput using more CPU.
Compared to Istio (Cilium), \name\ reduces 75.2\% (76.4\%) CPU utilization with a 60K req/s message rate.
When the message rate continues to increase, the throughput of Cilium becomes flat.
This is because its global proxy reaches its CPU quota, which is fixed internally by Cilium.
Unlike Cilium, Istio can get a throughput of around 100K req/s.
At this point, Istio achieves almost 100\% CPU utilization since its proxy has expensive per-core workers.
As a result, Istio cannot push higher throughput.
The highest throughput of \name\ is close to 300K req/s, which reaches to highest sending rate of 200 connections.
This is confirmed by that \name\ at most uses 75\% CPU.
To validate this, we keep adding more connections, and \name\ can support thousands of connections and achieve up to 600K req/s.

\subsection{Scalability and Interference}
\label{ss:scale}

\begin{figure*}[!t]
  \centering
  \includegraphics[width=6.7in]{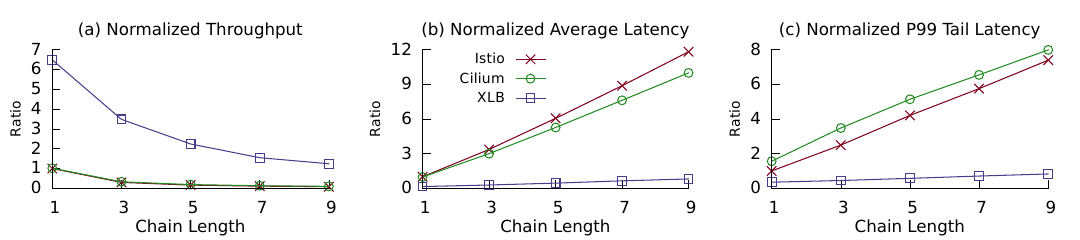}
  \caption{\small Performance of different chain length with zero size payload and 50 connections.
    All data are normalized to Istio one length result.
    }
  \label{fig:chain_len}
\end{figure*}

\begin{figure*}[!t]
  \centering
  \includegraphics[width=6.7in]{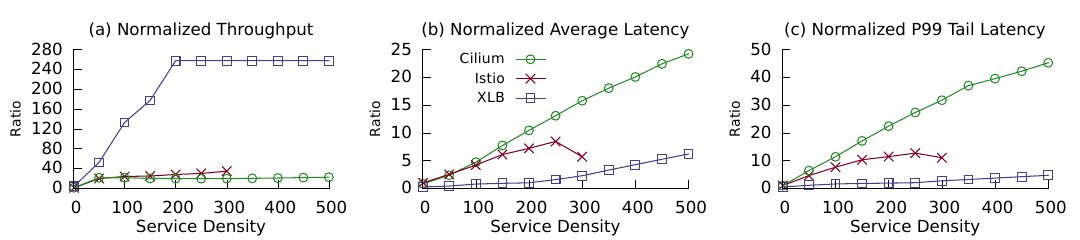}
  \caption{\small Performance results on an increasing number of services.
    Each service runs in a pod and has one connection.
    All data are normalized to Istio single service result.
    }
  \label{fig:instance}
\end{figure*}

Large scale microservices can hit many different scalability bottlenecks and suffer interference from concurrent services on the same host.
This section evaluates how \name\ helps scale out microservices while avoiding performance interference.

\head{Chain length.}
As pointed out by prior research, the performance of service chains is critical~\cite{edgeos_atc_20,audi_nsdi_21,spright_sigcom_22,nfvnice_sigcom_17}.
As discussed in \S\ref{s:intro}, the service chain imposes more challenges on L7 LB as they impact each hop.
Figure~\ref{fig:chain_len} depicts the performance results of running an increasing number of services in a chain.
As expected, the latency of a request grows linearly with the chain length in all systems.
\name\ has the lowest latency in all cases since it avoids per-hop proxy overhead.
For example, when the chain length is 9, \name\ has 93.1\% and 92.9\% lower latency than Istio and Cilium, respectively.
Because HTTP/1.1 does not allow concurrent requests, each client connection has to issue requests sequentially.
As a result, as the chain length and latency increase, the sending rate of each connection gets slower.
This is why the throughput of systems goes down.
Again, \name\ always maintains the highest throughput.

\begin{figure*}[!t]
  \centering
  \includegraphics[width=6.9in]{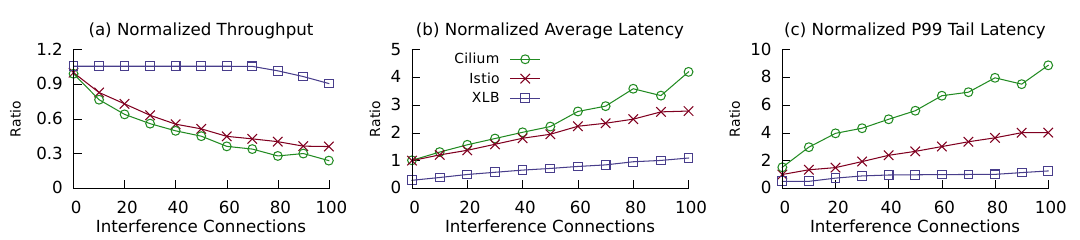}
  \caption{\small Performance results under increasing interference degree.
    The monitored service issues 5 connections with a fixed throughput of 5000 reqs/s.
    All data are normalized to Istio zero interference result.
    }
  \label{fig:interference}
\end{figure*}

\head{Service density.}
It is common that a single microservice application contains a large number of services because this provides great modularity, maintainability, and flexibility.
Additionally, cloud vendors also desire to consolidate as many as possible services on one host~\cite{lightvm_sosp_17,catalyzer_asplos_20,edgeos_atc_20}.
However, the proxies in existing LB can be a bottleneck.
We study this impact by booting up lots of client services in one VM and fixing the server to have 40 services running in another VM.
The results are shown in Figure~\ref{fig:instance}.
As the number of clients increases, the throughput goes up, but the latency also increases.
Around 100 services, Istio saturates the CPU, so its throughput becomes flat.
However, because Istio instantiates a per-service proxy, proxies quickly use up all resources of the VM.
Therefore, Istio can provision at most 300 services.
With 300 services, the execution of Istio is already unstable, and its reported results fluctuate greatly as shown in Figure~\ref{fig:instance}(b).
On the other hand, Cilium only has one global proxy, thus it can boot up more services.
However, the throughput of Cilium is again limited by its proxy capacity, and it achieves lower peak throughput than Istio.
Finally, \name\ outperforms the other systems.
\name\ saturates the CPU with around 200 services.
It still boots up more services, but the latency begins to go up due to queuing delay.

\begin{figure*}[!t]
  \centering
  \includegraphics[width=6.9in]{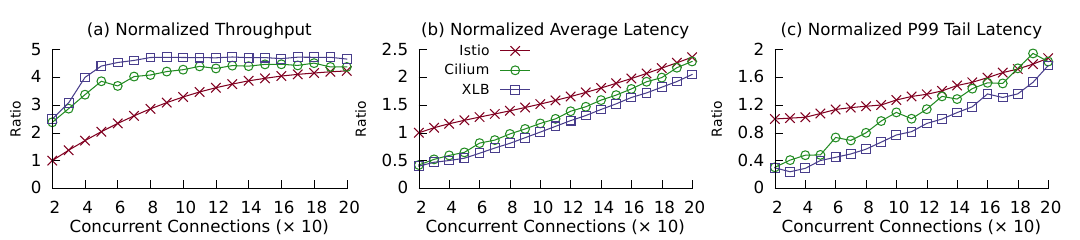}
  \caption{\small Bookinfo performance results.
    All data are normalized to Istio 20 connections result.
}
  \label{fig:bookinfo}
\end{figure*}

\head{Performance interference.}
QoS guarantee is extremely important for customers.
However, it is challenging to strictly ensure QoS in microservices due to the huge difference in traffic volume of each service.
For a specific service, there are three sources of interference.
\begin{inparaenum}[(1)]
\item Concurrent services will compete it with computing resources.
\item Proxies will also cause resource contention.
\item If the service shares a proxy with others, processing other services inside the proxy will disturb it as well.
\end{inparaenum}
To evaluate the degree of interference, we co-locate two client services on the same host.
The monitored client issues requests at a fixed rate, while the other client increases connections to generate interference.
Figure~\ref{fig:interference} shows the performance degradation of the monitored client under increasing interference.
Since Cilium has a single proxy shared by both clients, all the interference sources listed above apply to it.
Therefore, Cilium suffers the most interference and has the worst performance.
In Istio, each service has a separate proxy, thus Istio gets less interference and performs better than Cilium.
However, proxies in Istio take lots of CPU, which compete with the monitored client.
As a consequence, Istio gets worse performance as the interference increases.
\name\ is almost unaffected by the interference leading to the best performance.
\name\ handles every request efficiently and reserves enough CPU for the monitored client.
Furthermore, \name\ embeds the load-balancing logic into the in-kernel data path which exclusively processes the calling connection, thus avoiding contention with others.
Therefore, the only source of interference comes from the other client.
As shown in Figure~\ref{fig:interference}, when the other client issues many connections, the performance of \name\ begins to drop.

\subsection{Microservice Application}
\label{ss:micro_app}

\begin{figure*}[!t]
  \centering
  \includegraphics[width=6.85in]{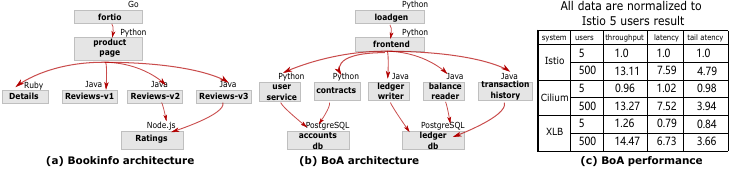}
  \caption{\small Microservice application architectures and performance results.
  }
  \label{fig:service}
\end{figure*}

To assess the practicability and applicability of \name, we evaluate \name\ under complicated microservice applications.
The first application is bookinfo~\cite{bookinfo_git}, a sample application from Istio.
Figure~\ref{fig:service}(a) shows its architecture.
We deploy it on two physical servers, and scale {\tt productpage} to 50 instances while other services have 5 instances.
Figure~\ref{fig:bookinfo} depicts the performance results of running an increasing number of client connections.
Since more time is spent in the service processing than previous micro-benchmarks, L7 LB have less impact on the performance.
However, \name\ still performs the best.
For instance, with 10 connections, \name\ has 43.2\% higher throughput, 33.2\% lower latency, and 39.6\% lower tail latency than Istio.
Similarly, \name\ has 10.2\% higher throughput, 13.3\% lower latency, and 21.2\% lower tail latency than Cilium.
Another application is Bank of Anthos (BoA) from Google~\cite{boa_git} (Figure~\ref{fig:service}(b)).
We run it using a 6 VM nods cluster.
We deploy 30 {\tt frontend} instances, 50 {\tt userservice} instances, and 5 instances for other services.
Figure~\ref{fig:service}(c) reports the normalized results.
With 500 useres, \name\ has 10.4\% higher throughput, 11.3\% lower latency and 23.5\% lower tail latency than Istio, 
9.1\% higher throughput, 10.6\% lower latency and 7.1\% lower tail latency than Cilium.
In summary, \name\ is more scalable than the current LB and can support large-scale microservice applications efficiently.

\section{Related Work}
\label{s:related}
\head{Load balancers.}
LB can operate on different network layers.
There is a variety of research improving L4 LB~\cite{anata_sigcom_13, duet_sigcom_14, rubik_atc_15, bypass_socc_20, maglev_nsdi_16, silkroad_sigcom_17, tiara_nsdi_22, pcc_nsdi_20, presto_sigcom_15, clove_conext_17} that are orthogonal to \name\ which focuses on L7.
Most of the existing research about L7 LB optimizes in-network LB that runs in a separate node.
For example, Yoda~\cite{yoda_eurosys_16} achieves high availability of L7 LB.
Prism~\cite{prism_socc_17, prism_nsdi_21} utilizes connection hand-off to accelerate content-routing proxies.
QDSR~\cite{298575} enables backend server instance to send data directly to the client to avoid extra message relaying.
Canal mesh~\cite{canal_sigcomm_24} deploys side-cars in a centralized mesh gateway, but still requires L7 traffic redirection to remote LBs. 
Some works offload LB to smart hardware for better performance~\cite{offload_apsys_21, appswitch_apnet_17}.
Different from the above works, \name\ targets microservices where LB is co-located with services in the same node, thus we can replace the sidecar-based architecture with in-kernel interposition.

\head{eBPF usage.}
eBPF is adopted in various Linux subsystems, such as scheduling~\cite{syrup_sosp_21, ghost_sosp_21}, memory prefetching~\cite{fetch_atc_24}, and storage~\cite{bpf_hotos_21,xrp_osdi_22,fuse_atc_19}, but is the most widely used in networking~\cite{deepflow_sigcomm_23,sketh_sigreview_23,inkv_sigreview_18,tango_nsdi_24,dctcp_nsdi_24}.
XDP~\cite{xdp_conext_18} hooks eBPF programs in the NIC driver to accelerate networking I/O, and hXDP~\cite{hxdp_osdi_20} further enables XDP programs to run on FPGA NICs.
BMC~\cite{bmc_nsdi_21} exploits eBPF to create an in-kernel memcached cache.
Electrode~\cite{eletrode_nsdi_23} leverages eBPF to implement in-kernel acceleration of distributed consensus protocols.
Packet steering is another use case that benefits from eBPF~\cite{steer_encp_19}.
DINT~\cite{dint_nsdi_24} offloads frequent-path transaction operations into eBPF to optimize distributed transaction systems.
{\tt bpf-iptables}~\cite{iptable_signcom_19} is an eBPF-based alternative to {\tt iptables}.
\cite{ebpf_survey} surveys challenges and applications of eBPF-based packet processing.
\name\ illustrates that in-kernel eBPF program is an efficient way to redesign application-level networking components, and as far as we know, \name\ is the first eBPF-based L7 LB.

\head{Serverless and microservice.}
Multiple projects optimize different aspects of serverless~\cite{sequoia_socc_20, sand_atc_18} and microservice frameworks~\cite{crisp_atc_22, nightcore_asplos_21}.
\name\ particularly optimizes load balancing for microservices.
SPRIGHT~\cite{spright_sigcom_22} combines eBPF-based event-driven processing and shared memory to improve the data plane of serverless function chains.
\name\ differentiates from SPRIGHT in two aspects.
First, SPRIGHT only optimizes the data path between functions in the same node, while \name\ accelerates message distribution between services that can cross multiple nodes.
Second, eBPF in SPRIGHT is simply used for basic message redirection without complex processing policies, but \name\ uses eBPF to implement richer routing and load balancing logic.
The eBPF design in SPRIGHT cannot be directly applied to implement and accelerate L7 redirection without the socket subtype system to parse and use L7 information or remove the overhead.

\section{Conclusions}
\label{s:conc}
The ubiquitous microservices present new challenges to efficient L7 LB.
The sidecar-based model becomes the main performance bottleneck due to its unavoidable overhead of system scheduling, process communication, and data movement.
This paper has described \name, an in-kernel L7 LB architecture that utilizes eBPF and extends socket to support load balancing, thus avoiding interweaving with underneath networking stacks.
\name\ is compatible with the Envoy control plane and load balancing rules.
Our evaluation shows \name\ achieves superior performance than Istio and Cilium.
With request-level load balancing over an application of 50+ instances, \name\ provides up to 1.5$\times$ (23\%) throughput increase, 60\% (12\%) reduction in average latency, and 70\% (30\%) lower tail latency than Istio (Cilium).
Furthermore,\name\ supports 66\% more service pods running on a host and achieves 13.6$\times$ higher throughput in a 9-services chain than Istio.

\bibliographystyle{plain}
\bibliography{xlb}

\end{document}